\documentclass{gretsi}

 \usepackage[english,french]{babel}   
  \usepackage{times}			
 \usepackage{url}
 \usepackage{graphicx}
 
\usepackage{multirow} 
 
\usepackage{xcolor,colortbl}

\definecolor{Gray}{gray}{0.85}
\definecolor{LightCyan}{rgb}{0.88,1,1}
\definecolor{DarkYellow}{rgb}{0.88,1,1}
\definecolor{LightYellow}{rgb}{0.88,1,1}
\definecolor{chartreuse(traditional)}{rgb}{0.87, 1.0, 0.0}

\newcolumntype{a}{>{\columncolor{Gray}}c}
\newcolumntype{b}{>{\columncolor{white}}c}

\newcolumntype{a}{>{\columncolor{red}}c}
\newcolumntype{b}{>{\columncolor{orange}}c}
\newcolumntype{d}{>{\columncolor{yellow}}c}
\newcolumntype{e}{>{\columncolor{chartreuse(traditional)}}c}
\newcolumntype{f}{>{\columncolor{green}}c}
 
 \usepackage[T1]{fontenc}

\titre{Comment partitionner automatiquement des marches al\'eatoires ?\\ Avec application \`a la finance quantitative}

\auteur{\coord{Gautier}{Marti}{1,2},
        \coord{Frank}{Nielsen}{2},
    \coord{Philippe}{Very}{1},
    \coord{Philippe}{Donnat}{1}}

\adresse{\affil{1}{Hellebore Capital Management \\
         63, avenue des Champs-Elys\'ees, 75008 Paris, France}
         \affil{2}{Laboratoire d'Informatique de l'Ecole Polytechnique \\
         1, rue Honor\'e d'Estienne d'Orves, 91120 Palaiseau, France}}

\email{gautier.marti@polytechnique.edu,
nielsen@lix.polytechnique.fr\\
philippe.very@helleborecapital.com,
philippe.donnat@helleborecapital.com}

\resumefrancais{Nous pr\'esentons dans cette communication une approche non param\'etrique pour regrouper automatiquement des s\'eries temporelles suivant une marche al\'eatoire. Nous introduisons d'abord une \'etape de pr\'e-traitement qui consiste \`a transformer les r\'ealisations ind\'ependantes et identiquement distribu\'ees des incr\'ements du processus de Markov en un vecteur repr\'esentant sans perte toute l'information disponible de ces s\'eries temporelles, et la factorisant en une composante d\'ependance et une composante distribution. Nous d\'efinissons ensuite une distance entre ces repr\'esentations tenant compte des deux types d'information et permettant d'en controler l'importance pour le partitionnement automatique \`a l'aide d'un seul param\`etre. 
Ce param\`etre de m\'elange peut \^etre appris ou manipul\'e par un expert \`a des fins exploratoires comme illustr\'e par l'\'etude des s\'eries temporelles financi\`eres. Des exp\'eriences, impl\'ementations et r\'esultats sont disponibles sur \url{http://www.datagrapple.com}.}

\resumeanglais{We present in this paper a novel non-parametric approach useful for clustering Markov processes. We introduce a pre-processing step consisting in mapping multivariate independent and identically distributed samples from random variables to a generic non-parametric representation which factorizes dependency and marginal distribution apart without losing any. An associated metric is defined where the balance between random variables dependency and distribution information is controlled by a single parameter. This mixing parameter can be learned or played with by a practitioner, such use is illustrated on the case of clustering financial time series. Experiments, implementation and results obtained on public financial time series are online on a web portal \url{http://www.datagrapple.com}.}

\begin{document}
\maketitle

\section{Introduction}

\begin{figure}[ht!]
\vskip 0.2in
\begin{center}
\centerline{\includegraphics[width=0.48\columnwidth]{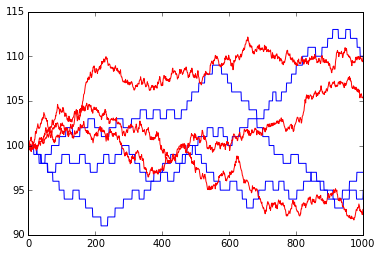}\includegraphics[width=0.48\columnwidth]{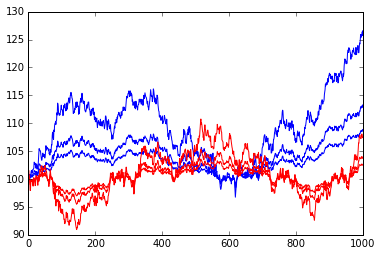}}
\caption{Pour regrouper ces exemples de marches al\'eatoires, deux crit\`eres sont utilis\'es: pour celles de gauche, la forme du signal; celles de droite sont similaires \`a transformations homoth\'etiques pr\`es.}
\label{random_walks}
\end{center}
\vskip -0.2in
\end{figure}

Les marches al\'eatoires peuvent \^etre utilis\'ees pour partitionner les donn\'ees, elles constituent par exemple un point de vue de la classification spectrale \cite{von2007tutorial}. Dans cette communication, nous nous int\'eresserons au probl\`eme inverse: partitionner des marches al\'eatoires. Ces processus stochastiques sont un important outil de mod\'elisation des s\'eries temporelles financi\`eres, savoir les regrouper dans des groupes homog\`enes statistiquement peut permettre d'\'etablir de meilleurs indicateurs de risque que la simple \og{}valeur \`a risque\fg{}.
Pour effectuer ce partitionnement automatique des marches al\'eatoires, nous devons disposer d'une repr\'esentation de celles-ci ainsi que d'une distance entre les repr\'esentations. En g\'en\'eral, repr\'esentation et distance idoines ne sont pas connues et des heuristiques sont utilis\'ees comme les deux d\'ecrites en l\'egende de la Figure~\ref{random_walks}. Dans le cas restreint des s\'eries temporelles s'\'ecrivant comme la somme $\sum_i X_i$ de variables al\'eatoires $X_i$ ind\'ependantes et identiquement distribu\'ees (i.i.d.), nous proposons en Section \ref{gnpr} distance et repr\'esentation adapt\'ees et math\'ematiquement fond\'ees. Celles-ci travaillent sur la s\'erie temporelle des incr\'ements $X_i$ portant toute l'information des marches al\'eatoires consid\'er\'ees. Finalement, en Section \ref{volume} nous pr\'esentons bri\`evement une application aux s\'eries temporelles financi\`eres. Pour une \'etude plus approfondie et davantage d'exp\'eriences, le lecteur pourra se r\'ef\'erer \`a \url{http://www.datagrapple.com}, portail se consacrant au partitionnement automatique des s\'eries temporelles, notamment issues du march\'e des couvertures de d\'efaillance.

\section{Une repr\'esentation non param\'etrique des marches al\'eatoires}\label{gnpr}

Soit $(\Omega,\mathcal{F},\mathbf{P})$ un espace de probabilit\'e.
Soit $\mathcal{V}$ l'espace des variables al\'eatoires r\'eelles continues d\'efinies sur $(\Omega,\mathcal{F},\mathbf{P})$.
Soient $\mathcal{U}$ l'espace des variables al\'eatoires suivant une loi uniforme sur $[0,1]$ et $\mathcal{G}$ l'espace des fonctions de r\'epartitions absolument continues. Nous d\'efinissons maintenant une repr{\'e}sentation non param\'etrique des vecteurs al\'eatoires qui capture et s\'epare sans perte la partie comportement joint des variables de leur distribution propre. Soit $\mathcal{T}$ l'application qui associe \`a un vecteur al\'eatoire $X = (X_1,\ldots,X_N)$ sa repr\'esentation non param\'etrique, \'el\'ement de $\mathcal{U}^N \times \mathcal{G}^N$, d\'efinit comme suit :
\begin{eqnarray}
\mathcal{T}:\mathcal{V}^N & \rightarrow & \mathcal{U}^N\times\mathcal{G}^N \\ 
X & \mapsto & (G_X(X),G_X) \nonumber
\end{eqnarray}
o\`u $G_X = (G_{X_1},\ldots,G_{X_N})$, $G_{X_i}$ \'etant la fonction de r\'epartition de $X_i$.

\begin{figure*}
\vskip 0.2in
\begin{center}
\centerline{\includegraphics[width=0.3\textwidth]{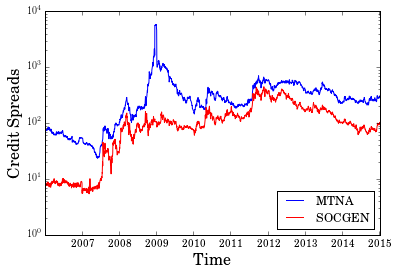}
\includegraphics[width=0.03\textwidth]{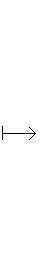}
\includegraphics[width=0.3\textwidth]{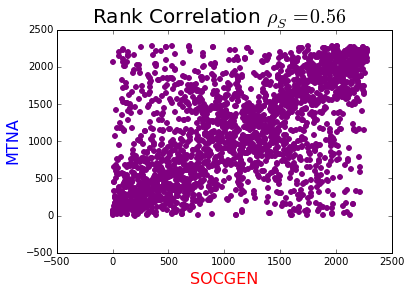} \includegraphics[width=0.03\textwidth]{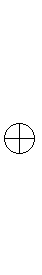} 
\includegraphics[width=0.29\textwidth]{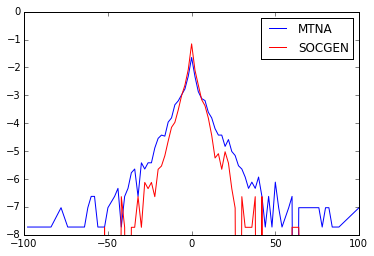}
}
\caption{L'approche pr\'esent\'ee en r\'esum\'e: deux s\'eries temporelles sont projet\'ees sur l'espace d\'ependance $\oplus$ distribution.}
\label{GNPR_projection}
\end{center}
\vskip -0.2in
\end{figure*}

$\mathcal{T}$ est une bijection et ainsi pr\'eserve la totalit\'e de l'information. La Figure~\ref{GNPR_projection} illustre cette projection sur un exemple concret issu de la finance. On peut remarquer que ce r\'esultat r\'eplique le th\'eor\`eme de Sklar \cite{sklar59}, r\'esultat fondateur de la th\'eorie des copules. N\'eanmoins, nous n'utilisons pas ici le cadre g\'en\'erique de cette th\'eorie et nous verrons par la suite o\`u cette analogie s'arr\^ete.
Nous exploitons ensuite cette repr\'esentation pour d\'efinir une distance $d_\theta$ entre les variables al\'eatoires qui prend en compte \`a la fois la distribution des marginales et leur comportement joint.

Soit $(X,Y) \in \mathcal{V}^2$. Soient $G_{X},G_{Y}$ leur fonction de r\'epartition.
Nous d\'efinissons la distance suivante, d\'ependante du param\`etre $\theta \in [0,1]$:
\begin{eqnarray}
d_{\theta}^{2}(X,Y)=\theta d_{1}^{2}(G_{X}(X),G_{Y}(Y))+(1-\theta)d_{0}^{2}(G_{X},G_{Y}), \nonumber
\end{eqnarray}
avec
\begin{eqnarray}
d_{1}^{2}(G_X(X),G_Y(Y))=3\mathbf{E}[\vert G_{X}(X)-G_{Y}(Y) \vert^{2}],
\end{eqnarray}
et
\begin{eqnarray}
d_{0}^{2}(G_{X},G_{Y})=\frac{1}{2}\int_{\mathbf{R}} \left(\sqrt{\frac{dG_{X}}{d\lambda}}-\sqrt{\frac{dG_{Y}}{d\lambda}}\right)^2 \, \mathrm{d}\lambda.
\end{eqnarray}

En particulier, nous obtenons $d_0$ la distance d'Hellinger, $f$-divergence qui quantifie la similarit\'e entre deux distributions et qui garantit la monotonicit\'e de l'information, propri\'et\'e qui assure que la distance entre des histogrammes grossiers est moindre que la distance entre des histogrammes plus pr\'ecis ; $d_1 = \sqrt{(1 - \rho_S) / 2}$ est une distance de corr\'elation mesurant la d\'ependance statistique entre deux variables al\'eatoires \`a l'aide de $\rho_S$, corr\'elation de Spearman entre $X$ et $Y$.
Remarquons que pour $\theta \in [0,1]$, $0 \leq d_\theta \leq 1$ et pour 
$0 < \theta < 1$, $d_\theta$ est une distance m\'etrique. Pour $\theta = 0$ ou $\theta = 1$, l'axiome de s\'eparation n'est pas v\'erifi\'e.
Cette distance est \'egalement invariante par transformations monotones, propri\'et\'e d\'esirable car elle affranchit de l'arbitraire du choix des unit\'es ou de la m\'ethode de mesure (que ce soit l'appareillage ou la mod\'elisation math\'ematique) du signal.

Pour appliquer la distance propos\'ee sur des donn\'ees \'echantillonn\'ees, nous d\'efinissons alors une estimation de $d_\theta$. La distance $d_1$ travaillant avec des distributions uniformes continues peut \^etre approxim\'ee de mani\`ere discr\`ete par des statistiques de rang qui en sus d'\^etre robustes aboutissent \`a une analogie avec le formalisme des copules: la statistique de rang utilis\'ee correspond \`a une coordonn\'ee de la copule empirique de Deheuvels \cite{deheuvels79} qui est un estimateur non param\'etrique et non biais\'e convergeant uniform\'ement \cite{deheuvels81} vers la copule sous-jacente au processus.
La distance $d_0$ peut \^etre approxim\'ee par sa forme discr\`ete travaillant sur une estimation des densit\'es marginales obtenues par histogrammes, par exemple.
Pour calculer $d_1$, nous avons besoin d'une fonction de rang bijective et puisque nous consid\'erons l'application aux s\'eries temporelles, il est naturel de privil\'egier l'ordre d'arriv\'ee pour d\'epartager les \'egalit\'es.

Soient $(X_i)_{i=1}^M$ les $M$ r\'ealisations de $X \in \mathcal{V}$.
Soit $S_M$ le groupe des permutations de $\{1,\ldots,M\}$ et $\sigma \in S_M$ une permutation quelconque, disons $\sigma = Id_{\{1,\ldots,M\}}$. Une fonction de rang bijective pour $(X_i)_{i=1}^M$ peut \^etre d\'efinie comme une fonction
\begin{eqnarray}
\mathrm{rk}^X : \{1,\ldots,M\} & \rightarrow & \{1,\ldots,M\} \\
i & \mapsto & \# \{ k \in \{1,\ldots,M\} ~|~ \mathcal{P}_\sigma \} \nonumber
\end{eqnarray}
avec $\mathcal{P}_\sigma \equiv (X_k < X_i) \lor (X_k = X_i \land \sigma(k) \leq \sigma(i))$.

Soient $(X_i)_{i=1}^M$ et $(Y_i)_{i=1}^M$ les $M$ r\'ealisations des variables al\'eatoires $X,Y \in \mathcal{V}$.
Une distance empirique entre les r\'ealisations de ces variables al\'eatoires peut \^etre d\'efinie par
\begin{eqnarray}
\tilde{d}_\theta^2\left((X_i)_{i=1}^M,(Y_i)_{i=1}^M\right) \stackrel{a.s.}{=} \theta \tilde{d}_{1}^2 + (1-\theta) \tilde{d}_{0}^2,
\end{eqnarray}
avec
\begin{eqnarray}
\tilde{d}_1^2 = \frac{3}{M^2(M-1)}\sum_{i=1}^M \left(\mathrm{rk}^X(i) - \mathrm{rk}^Y(i)\right)^2
\end{eqnarray}
et
\begin{eqnarray}
\tilde{d}_0^2=\frac{1}{2}\sum_{k=-\infty}^{+\infty} \left(\sqrt{g_X^h(hk)} - \sqrt{g_Y^h(hk)} \right)^2,
\end{eqnarray}
le param\`etre $h$ \'etant un param\`etre de lissage appropri\'e, et $g_X^h(x) = \frac{1}{M}\sum_{i=1}^M\mathbf{1}\{ \lfloor \frac{x}{h} \rfloor h\leq X_i < (\lfloor \frac{x}{h} \rfloor +1)h\}$ \'etant un histogramme de densit\'e estimant la fonction de densit\'e de probabilit\'e $g_X$ \`a partir des $(X_i)_{i=1}^M$, les $M$ r\'ealisations de la variable al\'eatoire $X \in \mathcal{V}$.

\section{Application au partitionnement automatique de s{\'e}ries temporelles financi{\`e}res}\label{volume}

\begin{figure}[ht!]
\vskip 0.2in
\begin{center}
\centerline{\includegraphics[width=\columnwidth]{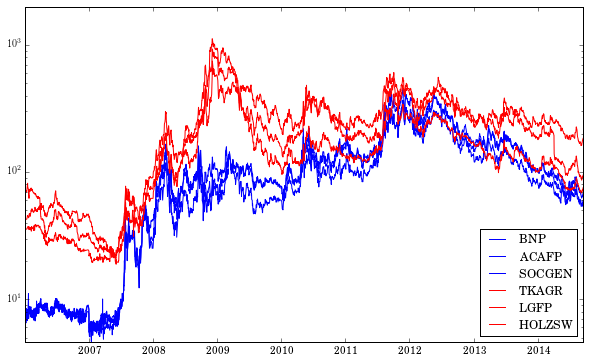}}
\caption{Les prix de CDS de deux industries entre janvier 2006 et janvier 2015 : les entreprises financi\`eres fran\c caises (en bleu) et les cimentiers (en rouge) ; observez la corr\'elation importante \`a l'int\'erieur de chaque secteur industriel.}
\label{CDS_spreads}
\end{center}
\vskip -0.2in
\end{figure}

\begin{figure}[!t]
\centering
\includegraphics[width=\linewidth]{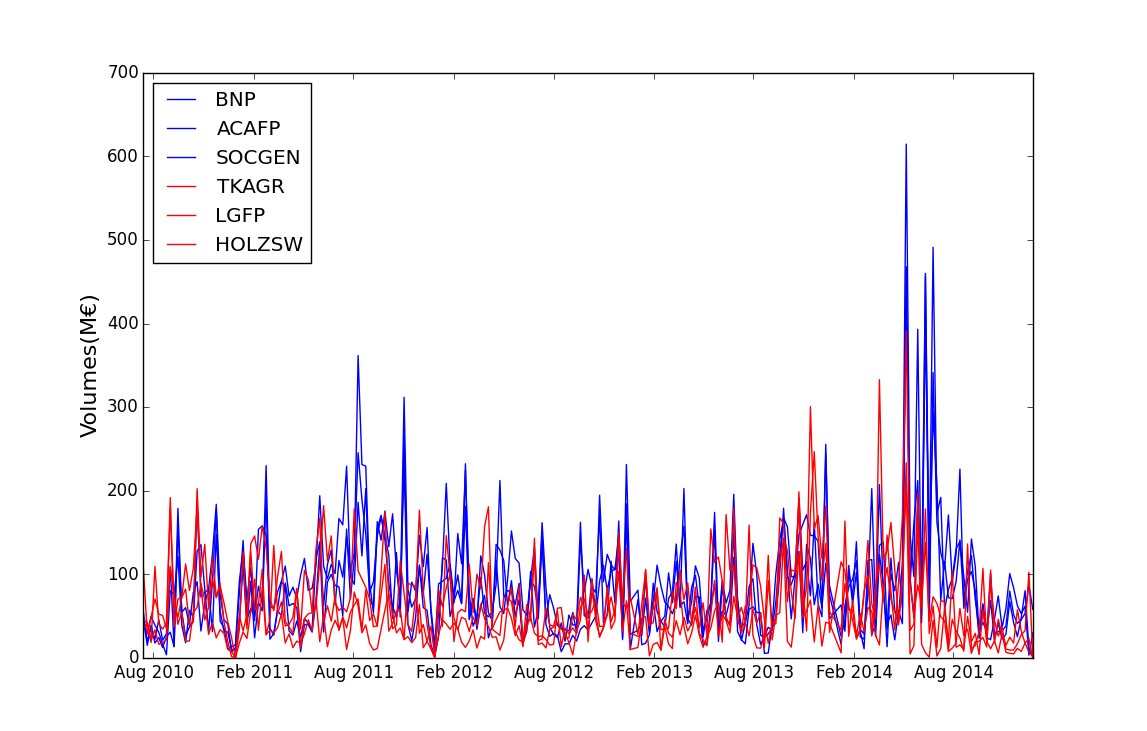}
\caption{Les volumes de CDS trait\'es selon DTCC; En bleu, les entreprises financi\`eres fran\c caises et en rouge les volumes trait\'es sur les cimentiers tels que report\'es entre juillet 2010 et janvier 2015.}\label{fig:dtcc_volumes}
\end{figure}

Nous illustrons notre approche sur les s\'eries temporelles des volumes trait\'es sur le march\'e des couvertures de d\'efaillance \cite{hull06} (CDS).
Nous prenons en compte les $N = 658$ actifs ayant des volumes report\'es depuis juillet 2010. En sus d'\^etre des donn\'ees accessibles publiquement (fournies par DTCC - \url{http://www.dtcc.com/}) contrairement aux prix des CDS, ces s\'eries temporelles sont tr\`es bruit\'ees et font montre de moins de corr\'elations \'evidentes que les s\'eries de prix \cite{kane11} (cf. Figure~\ref{CDS_spreads} et Figure~\ref{fig:dtcc_volumes} pour une comparaison), ce qui rend ce jeu de donn\'ees int\'eressant pour notre m\'ethode. A notre connaissance, il s'agit de la premi\`ere fois qu'un papier s'int\'eresse au regroupement automatique de s\'eries temporelles des volumes trait\'es sur un march\'e financier.

Notre but est de comprendre comment ces s\'eries temporelles se regroupent lorsque nous consid\'erons uniquement leur comportement joint (notre approche avec $\theta = 1$) ou en se concentrant seulement sur la proximit\'e de la distribution de leurs volumes trait\'es (notre approche avec $\theta = 0$), et finalement lorsque nous prenons en compte la totalit\'e de l'information (notre approche avec $\theta = 0.5$). Nous estimons d'abord le nombre de groupes dans chaque cas gr\^ace \`a un crit\`ere de stabilit\'e \cite{shamir2008model} et nous trouvons $K_1 = 3$, $K_0 = 5$ et $K_{0.5} = 7$ respectivement.

La Table \ref{tab3} affiche quelques caract\'eristiques (esp\'erance et quantiles) de la distribution des $K_{0.5} = 7$ groupes trouv\'es en utilisant la totalit\'e de l'information. Nous pouvons remarquer que ces groupes correspondent en fait aux $K_0 = 5$ groupes trouv\'es en utilisant uniquement l'information de distribution dont les esp\'erances et quantiles sont report\'es dans la Table \ref{tab4}. Cependant, ces indicateurs sur la distributions ne permettent pas d'expliquer les diff\'erences entre les groupe 3 et 4 qui se ressemblent pour ces mesures, idem pour les groupes 5 et 6.

Concernant $\{C_1^{0.5},C_2^{0.5},C_7^{0.5}\}$, nous pouvons d'ores et d\'ej\`a constater que $C_1^{0.5}$ est compos\'e des CDS ayant un important volume trait\'e, notamment les CDS sur la dette souveraine de pays tels que le Br\'esil, la Chine, l'Allemagne, la France, l'Italie, la Russie et l'Espagne. $C_2^{0.5}$ est constitu\'e des entreprises financi\`eres ainsi que de quelques fournisseurs d'\'energie qui repr\'esentent les entit\'es les plus activement trait\'ees sur le march\'e des couvertures de d\'efaillance, en dehors des dettes souveraines.
$C_7^{0.5}$ se compose des entreprises asiatiques, notamment japonaises, dont les CDS sont relativement peu trait\'es, les rendements \'etant tr\`es faibles. Pour comprendre les diff\'erences entre les groupes $C_3^{0.5}$, $C_4^{0.5}$ et $C_5^{0.5}$, $C_6^{0.5}$, nous \'etudions les r\'esultats du regroupement automatique en utilisant seulement les comportements joints, c'est-\`a-dire les $K_1 = 3$ groupes $\{C_1^1,C_2^1,C_3^1\}$. $C_1^1$ est essentiellement compos\'e d'entit\'es ayant une liquidit\'e croissante, c'est-\`a-dire une tendance haussi\`ere des volumes trait\'es, et correspond au groupe $C_6^{0.5}$. $C_2^1$ contient les CDS des entreprises europ\'eennes consid\'er\'ees comme \'etant s\^ures par les agences de notations, ce march\'e est connu pour \^etre tr\`es fortement corr\'el\'e en comparaison de ses \'equivalents am\'ericain et asiatique. $C_3^1$ semble rassembler le reste des actifs ne partageant pas de points communs \'evidents.
\begin{table}[!t]
\caption{\label{tab3}Les $K_{0.5} = 7$ groupes obtenus avec $\theta = 0.5$}
\centering
\begin{small}
\begin{tabular}{|c|a|b|d|d|e|e|f|}
\hline
 & C$_1^{0.5}$ & C$_2^{0.5}$ & C$_3^{0.5}$ & C$_4^{0.5}$ & C$_5^{0.5}$ & C$_6^{0.5}$ & C$_7^{0.5}$ \\
\hline
\multirow{1}{*}{Mean} &
441 & 84 & 32 & 29 & 17 & 17 & 8 \\
\hline
\multirow{1}{*}{Quantile 10\%}
 &
116 & 46 & 18 & 17 & 8 & 5 & 4 \\
\hline
\multirow{1}{*}{Quantile 90\%}
 &
924 & 141 & 50 & 44 & 29 & 36 & 15 \\
\hline
\multirow{1}{*}{Size}
 & 13 & 89 & 169 & 79 & 161 & 90 & 57 \\
\hline
\multicolumn{6}{@{}l}{}
\end{tabular}
\end{small}
\end{table}

\begin{figure}[ht!]
\vskip 0.2in
\begin{center}
\centerline{\includegraphics[width=\columnwidth]{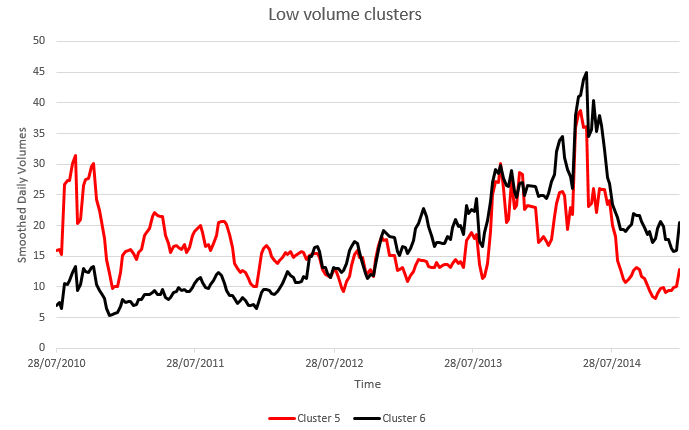}}
\caption{Des dynamiques inverses pour $C_5^{0.5}$ et $C_6^{0.5}$}
\label{TimeStab_centers56}
\end{center}
\vskip -0.2in
\end{figure}

\begin{table}[!t]
\caption{\label{tab4}Les $K_{0} = 5$ groupes obtenus avec $\theta = 1$}
\centering
\begin{tabular}{|c|a|b|d|e|f|}
\hline
 & C$_1^{0}$ & C$_2^{0}$ & C$_3^{0}$ & C$_4^{0}$ & C$_5^{0}$\\
\hline
\multirow{1}{*}{Mean} &
458 & 92 & 40 & 22 & 10\\
\hline
\multirow{1}{*}{Quantile 10\%}
 &
196 & 60 & 29 & 16 & 4\\
\hline
\multirow{1}{*}{Quantile 90\%}
 &
924 & 139 & 51 & 29 & 15\\
\hline
\multicolumn{6}{@{}l}{}
\end{tabular}
\end{table}
Nous pensons que ces volumes trait\'es constituent un jeu de donn\'ees int\'eressant pour illustrer l'usage de notre m\'ethode car cela montre le gain qu'on obtient \`a exploiter l'information totale disponible dans ces marches al\'eatoires. En sus, nous trouvons que le regroupement automatique optimal (d'un point de vue de la stabilit\'e des groupes par rapport \`a des petites perturbations) est constitu\'e des groupes qui sont eux-m\^emes r\'esultats optimaux des regroupements automatiques lorsque l'algorithme travaille seulement sur la partie \og{}d\'ependance\fg{} de l'information ou seulement sur la partie \og{}distribution\fg{}: les CDS sont regroup\'es en 5 groupes pouvant \^etre expliqu\'es par le volume moyen trait\'e et qui r\'esume approximativement l'information de distribution, cependant deux groupes suppl\'ementaires \'emergent \`a cause de l'information sur les comportements joints qui raffine cette partition en 5 groupes: un groupe \'emerge \`a cause des fortes corr\'elations pr\'esentes dans le march\'e europ\'een des actifs s\^urs, et l'autre rassemble les entit\'es dont le volume des transactions est en augmentation (Figure~\ref{TimeStab_centers56}).

\section{Discussion}

Dans cette communication, nous avons pr\'esent\'e une nouvelle repr\'esentation, math\'ematiquement fond\'ee, des s\'eries temporelles suivant une marche al\'eatoire. Cette repr\'esentation peut \^etre utilis\'ee pour le partitionnement automatique des s\'eries temporelles comme illustr\'e en Section \ref{volume} par l'exemple des volumes trait\'es, mais est \'egalement adapt\'ee \`a l'apprentissage supervis\'e. Dans cette communication, nous avons montr\'e son utilit\'e sur des donn\'ees r\'eelles, n\'eanmoins nous avons \'egalement valid\'e l'approche sur des cas tests engendr\'es par des mod\`eles de corr\'elations hi\'erarchiques se subdivisant en groupes de distribution. Nous nous concentrons maintenant \`a prouver la consistance statistique d'une telle approche. Les r\'esultats exp\'erimentaux, des donn\'ees ainsi que des impl\'ementations, sont disponibles sur \url{http://www.datagrapple.com} se consacrant au partitionnement automatique de s\'eries temporelles.

\section*{Remerciements}

Merci \`a Valentin Geffrier et Benjamin d'Hayer pour leur relecture attentive, et Laurent Beruti pour son retour et son expertise sur le march\'e des CDS.


\begin{thebibliography}{99}

\bibitem{deheuvels79}
P.~Deheuvels.
\emph{La fonction de d\'ependance empirique et ses propri\'et\'es. Un test non param\'etrique d'ind\'ependance}.
Acad. Roy. Belg. Bull. Cl. Sci.(5), 1979.

\bibitem{deheuvels81}
P.~Deheuvels.
\emph{An asymptotic decomposition for multivariate distribution-free tests of independence}.
Journal of Multivariate Analysis, 1981.

\bibitem{hull06}
J.~Hull.
\emph{Options, futures, and other derivatives}.
Pearson Education, 2006.

\bibitem{kane11}
D.~Kane.
\emph{Modelling single-name and multi-name credit derivatives}.
John Wiley \& Sons, 2011.

\bibitem{shamir2008model}
O.~Shamir et T.~Naftali.
\emph{Model selection and stability in k-means clustering}.
Conference on Learning Theory, 2008.

\bibitem{sklar59}
A.~Sklar.
\emph{Fonctions de r\'epartition \`a n dimensions et leurs marges}.
Universit\'e Paris 8, 1959.

\bibitem{von2007tutorial}
U.~Von Luxburg.
\emph{A tutorial on spectral clustering}.
Statistics and computing, 2007.


\end{thebibliography}
\end{document}